# Electrochemical Deposition of FeSe on RABiTS Tapes

Satoshi Demura[1,2], Masashi Tanaka[1*], Aichi Yamashita[1,3], Saleem J. Denholme[1,2], Hiroyuki Okazaki[1], Masaya Fujioka[1], Takahide Yamaguchi[1], Hiroyuki Takeya[1], Kazumasa Iida[4,5], Bernhard Holzapfel[4], Hideaki Sakata[2], and Yoshihiko Takano[1,2,3]

[1]*National Institute for Materials Science, 1-2-1 Sengen, Tsukuba, Ibaraki 305-0047, Japan*

[2]*Tokyo University of Science, 1-3 Kagurazaka, Shinjyuku, Tokyo 162-8601, Japan*

[3] *Graduate School of Pure and Applied Sciences, University of Tsukuba, 1-1-4 Tennodai, Tsukuba, Ibaraki 305-8577, Japan*

[4]*Institute for Metallic Materials, IFW Dresden, D-01171 Dresden, Germany*

[5] *Department of Crystalline Materials Science, Nagoya University, Chikusa, Nagoya 464-8603, Japan*

*E-mail: Tanaka.Masashi@nims.go.jp

**Abstract**

FeSe film is successfully fabricated onto rolling assisted biaxially textured substrate (RABiTS) tapes by an electrochemical deposition technique. The deposited FeSe films tend to become high crystallinity with a decrease in the applied voltage to -1.0 V, and the compositional ratio of Fe to Se approaches 1:1. The sample deposited at -1.0 V shows a superconducting transition at approximately 8.0 K in the magnetic susceptibility.

After the discovery of superconductivity in iron-based compounds [1], much attention has been paid to the development of superconducting applications under a high magnetic field [2], because of their high upper critical and irreversibility fields [3]. The iron chalcogenides FeSe are one of the members of iron-based superconductors with the simplest structure among them. The original transition temperature $T_c$ of 8 K [4] increases above 30 – 46 K by the intercalation of alkali or alkaline earth elements [5-10]; it also increases to 37 K under high pressure [11-14]. The large increase in $T_c$ is attributed to the sensitivity in its local structural difference such as the anion height from Fe atoms [15]. This is related to the relatively higher $T_c$ of FeSe tapes or thin films than that of bulk FeSe because of its lattice strain [16-18]. These facts indicate that FeSe is a potential material for superconducting wires or tapes. It is important for industrial applications to develop simple and low-cost methods of fabricating thin films over large areas.

We have been developing the deposition of FeSe by the electrochemical technique. Superconducting FeSe films were deposited onto an Fe substrate by this method [19,20]. This is the first case of the preparing tape-shaped superconducting materials by an electrochemical method. If we can change the substrate to other metals, it would lead to a wide variety of the fabrication processes.

In this study, we chose the rolling assisted biaxially textured substrates (RABiTS) tape as a substrate, and performed an electrochemical deposition of FeSe onto the tape. The deposited FeSe shows superconductivity at approximately 8.0 K, comparable to the $T_c$ of bulk materials. This is part of the studies to develop fabricating conditions for FeSe as a thin film by an electrochemical technique.

The electrochemical depositions were performed for 30 min at a constant voltage by a three-electrode method. A platinum plate, a RABiTS tape (evico GmbH), and a Ag/AgCl electrode were used as an anode, a cathode, and a reference electrode, respectively. The RABiTS tape is composed of (001)-plane-oriented Ni formed by the recrystallization of cold-rolled pure Ni. A mixture of 0.03 mol/l $FeCl_2 \cdot 4H_2O$, 0.015 mol/l $SeO_2$, and 0.1 mol/l $Na_2SO_4$ was dissolved in distilled water for the preparation of the electrolyte. The pH of the electrolyte was adjusted to 2.1 by adding $H_2SO_4$. Powder X-ray diffraction (XRD) measurements were carried out using Mini Flex 600 (RIGAKU) with Cu-K$\alpha$ radiation. The compositional ratio was analyzed by energy dispersive X-ray spectroscopy using JSM-6010LA (JEOL). The temperature dependence of magnetic susceptibility was measured using a SQUID magnetometer (MPMS, Quantum Design) down to 2.0 K under a field of 100 Oe.

Cyclic voltammetry (CV) for the electrolyte was performed between 0.0 and -3.0 V against the reference electrode in order to identify a suitable voltage for the deposition of FeSe (Fig. 1(a)). An anomaly was observed at ~ -1.1 V in each voltammogram. A similar anomaly was also observed in the CV result of the previous paper, in which the FeSe was deposited on the Fe substrate [20]. Thus, the anomaly in the CV measurement is attributed to the electrodeposition of FeSe. Then, a voltage range between -0.8 and -1.1 V was employed for the deposition in this study.

Figure 1(b) shows the XRD patterns of the peeled off samples from the resulting substrates after the electrochemical deposition at various voltages. The crystallinity tends to become higher with the decrease in the applied voltage to -1.0 V. The sample obtained at a voltage of -1.0 V showed almost the single phase of FeSe with relatively sharpe peaks. The diffraction peaks can be indexed on the basis of a tetragonal unit cell with the lattice parameters $a$ = 3.785(6) Å and $c$ = 5.5076 (12) Å, which are in good agreement with those of reported one ($a$ = 3.77376(2) Å, $c$ = 5.52482(5) Å) [21]. The samples obtained at -0.8 and -0.9 V also showed almost the single phase of FeSe with the same lattice parameters within the error. On the other hand, no diffraction peaks from FeSe appeared at a voltage of -1.1 V, indicating that the voltage is too low to deposit the FeSe crystals.

The molar ratio of Fe and Se in the obtained samples is shown in Fig. 1(c). The values of Fe and Se were comparable at -1.1 V. However, the compositional ratio deviates from Fe:Se = 1:1 with increasing voltage, even though the results of XRD apparently showed almost the single phase. This suggests that the FeSe starts to deposit at around -0.8 V and increases the yield with decreasing voltage. This result is consistent with the result of the crystallinity enhancement in XRD patterns. In our previous study of FeSe deposition on the Fe substrate [20], the production of the FeSe film was very sensitive to the applied voltage; the diffraction peaks of the film immediately broadened when the voltage deviated from the appropriate voltage. The result at -1.1 V in this study is attributed to the voltage sensitivity of the deposition.

Figure 1(d) shows the temperature dependence of the magnetic susceptibility of the sample deposited at -1.0 V. The susceptibility showed a downturn at approximately 8.0 K in its zero-field cooling, indicating the superconducting transition. The positive signal in the whole region may be a contribution of simultaneously deposited Fe.

From all these results, a superconducting FeSe film was partially deposited on RABiTS tape by electrochemical deposition in the voltage range of -0.8 – -1.0V. This finding indicates that the superconducting wire can be fabricated by just passing through a specific liquid with applying voltage, if we can adjust the voltage appropriately. This method can lead to a tremendous reduction in the production cost for wire fabrication compared with the conventional way. Further investigations on the optimum synthesis conditions will lead to the formation of a perfect FeSe film with a high critical current, by modifying the concentration, temperature, pH of the electrolyte, the reaction time, and so on.

Acknowledgements

This work was partly supported by a Grant-in-Aid for Scientific Research from the Ministry of Education, Culture, Sports, Science and Technology (KAKENHI) and the Strategic International Collaborative Research Program (SICORP-EU-Japan).

**Figure caption**

Fig. 1(Color online) (a) Cyclic voltammogram in distilled water dissolving 0.03 mol/l $FeCl_2 \cdot 4H_2O$, 0.015 mol/l $SeO_2$, and 0.1 mol/l $Na_2SO_4$. The red and blue lines correspond to the forward and reverse scans, respectively. (b) X-ray diffraction patterns of FeSe deposited on RABiTS tape in a voltage range between -0.8 and -1.1 V. All the indices are given on the basis of the primitive tetragonal unit cell. (c) Voltage dependence of the molar ratio of Fe and Se. (d) Temperature dependence of magnetic susceptibility for the sample synthesized at -1.0 V in zero-field cooling (ZFC) and field cooling (FC) modes.

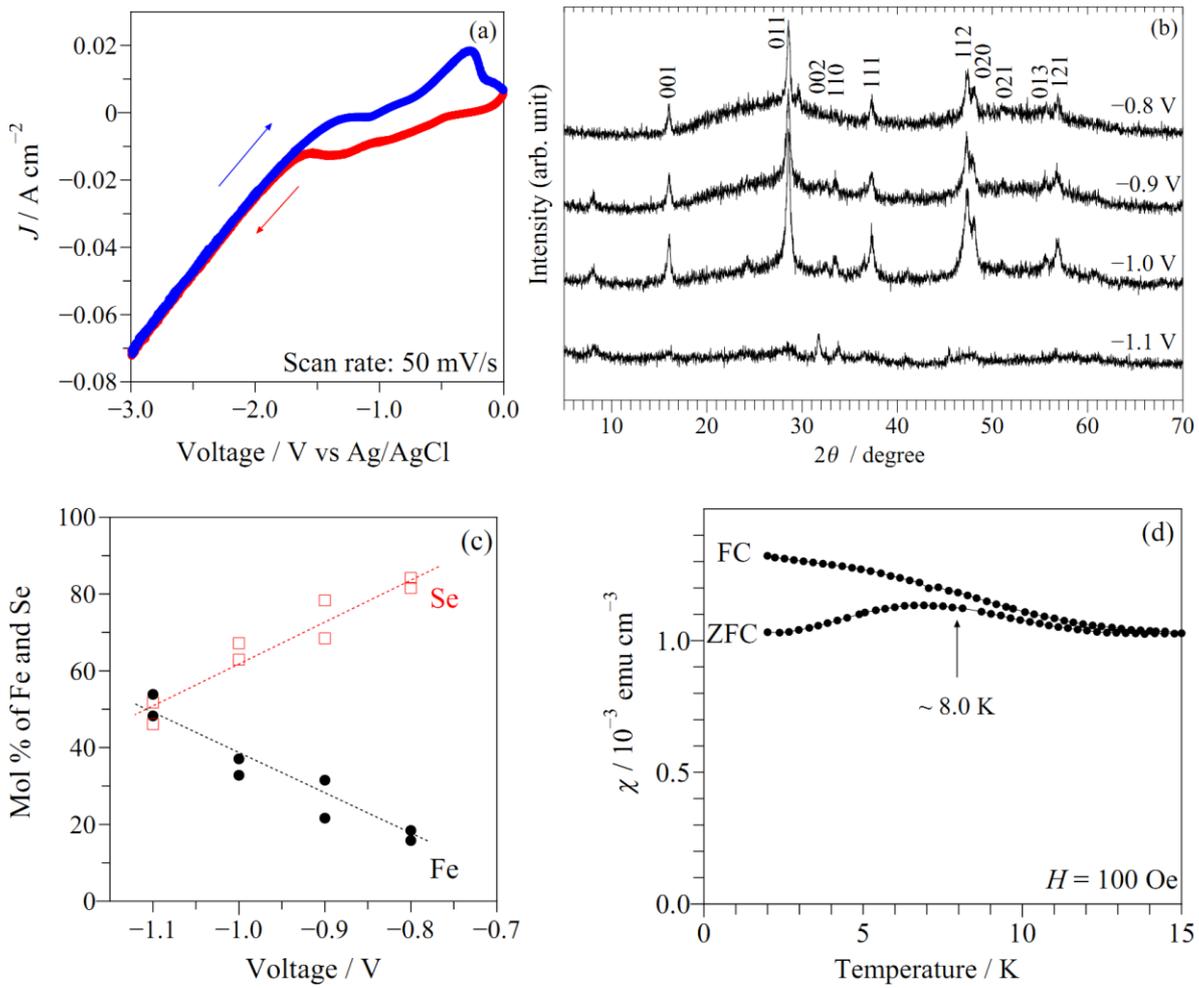

Fig.1